\documentclass{article}

\usepackage{amssymb}
\usepackage{amsmath} 
\usepackage{amsbsy} 
\usepackage{graphicx}

\begin{document}

\newcommand{\delh}{\delta\!h}
\newcommand{\delu}{\delta\!u}

\title{Spinodal dewetting of thin films with large 
interfacial slip: implications from the dispersion relation
}

\author{Markus Rauscher,$^{ab}$\footnote{To whom correspondence should be
addressed. E-mail: rauscher@mf.mpg.de}\ Ralf Blossey,$^{c}$
Andreas M\"unch,$^{d}$ \\and
Barbara Wagner$^{e}$                     \\[2ex]
$^a$Max-Planck-Institut f\"ur Metallforschung,\\
Heisenbergstr. 3, 70569 Stuttgart, Germany,\\[1ex]
$^b$ITAP, Universit{\"a}t Stuttgart, \\Pfaffenwaldring 57, 70569 Stuttgart,
Germany,\\[1ex]
$^c$Interdisciplinary Research Institute, c/o IEMN Avenue\\
Poincar\'e BP 60069, F-59652 Villeneuve d'Ascq, France \\ [1ex]
$^d$School of Mathematical Sciences, \\University of Nottingham, NG7 2RD,
UK \\ [1ex]
$^e$Weierstrass Institute for Applied Analysis and Stochastics,\\
Mohrenstrasse 39, 10117 Berlin, Germany
}

\date{Version of \today}

\maketitle

\abstract{
We compare the dispersion relations for spinodally dewetting thin
liquid films for increasing magnitude of interfacial slip length in
the lubrication limit. While the shape of the dispersion relation,
in particular the position of the maximum, are equal for no-slip up to
moderate slip lengths, the position of the maximum shifts to much
larger wavelengths for large slip lengths. Here, we discuss the
implications of this fact for recently developed methods to assess the
disjoining pressure in spinodally unstable thin films by measuring
the shape of the roughness power spectrum. For PS films on OTS
covered Si wafers (with slip length $b\approx 1\,\mu$m) we predict a
20\%\ shift of the position of the maximum of the power spectrum
which should be detectable in experiments.
}

\section{Introduction}
\label{sec:intro}

Wetting and dewetting phenomena are not only part of our everyday
life but they are particularly relevant to technological
applications (e.g., in coating processes) and in biological
systems. The dynamics of films of a thickness smaller than 10 or 20
nanometers is not only governed by hydrodynamics, but the finite
range of intermolecular forces, which are responsible for the
richness of wetting phenomena \cite{degennes85,dietrich88}, becomes
relevant \cite{oron97}\/. This is true in particular for spinodally
unstable films which have been analysed quantitatively in the
framework of the thin-film equation \cite{becker03}\/. In addition,
the quantitative analysis of he roughness power spectrum has been
used in order to measure the disjoining pressure (DJP) (or the effective
interface potential) between the liquid-solid and the liquid-vapor
interface which is a result of the interplay beween the
interactions among the fluid molecules and the interactions between
the fluid and the substrate
\cite{bischof96,seemann01d,seemann01b}\/.

In the spinodally dewetting systems studied in
Refs.~\cite{seemann01d,seemann01b,becker03}, i.e., polystyrene (PS) on
silicon (Si) wafers covered with a native oxide layer, hydrodynamic slip
between the fluid and the solid substrate could be neglected. However,
recently, the slip length of PS on octadecyltrichlorosilane (OTS) and 
decyltrichlorosilane (DTS) coated Si wafers was
discovered to range up to the scale of a micron
\cite{fetzer05,fetzer06a,fetzer07a,fetzer07c}\/.
The dewetting patterns, in particular the shape of the dewetting
rims around the growing holes in the film were analyzed using a
thin-film equation valid in the regime of large slip lengths, the so-called  
strong-slip model, \cite{fetzer05,witelski05}. The thickness of the films
was on the order of a few 100~nm and the dewetting mechanism was
therefore nucleation rather than spinodal. However, since the
hydrodynamic boundary conditions influence the rim shape it is to
be expected that the power spectrum of spinodally unstable films
is affected as well. Some dependencies of the 
dominant wavelength and time scale on the magnitude of the slip 
length for the case of an attracting van der Waals potential 
are discussed in \cite{kargupta04}\/. 
In this study, we systematically compare thin spinodally
dewetting films with
zero to large large slip lengths. Our main motivation
is the availability of experimental systems (PS on
OTS or DTS-covered Si wafers
\cite{fetzer05,fetzer06a,fetzer07a,fetzer07c})
which exhibit extremely large slip
lengths and against which we can test our
theoretical analysis in order to not only infer qualitative but
also quantitative results.
In particular, we consider an effective interface potential 
calculated from Hamaker constants 
as given in \cite{seemann01b}\/.

In the following Sec.~\ref{sec:dewet} we first compare the
dispersion relations for lubrication models for zero to moderate
slip lengths with the dispersion relation for the regime of large
slip lengths.  We establish that even though the energetics is the
same, films dewetting for large slip lengths have a qualitatively
different dispersion relation as compared to sticky films, and
therefore their structure factor has a maximum at a different
wavenumber. We investigate the relevance of the difference
between the dispersion relations for the sample systems PS on
Si and on OTS(DTS)-covered Si in Sec.~\ref{sec:discussion} and
conclude in Sec.~\ref{sec:summary}.

For clarity of presentation we restrict
our analysis to one-dimensional interfaces. The generalization 
to real two-dimensional interfaces is straightforward for the 
lubrication models for zero to moderate slip.
The generalization of the lubrication model for large slip is not
completely obvious due to the appearance of additional cross-terms
and we have included it here in an appendix \cite{ambw08}\/.
We note that omission of these cross-terms would be discovered
at the level of the dispersion relation, where the growth rate would
not only depend on the modulus of the wave vector but also on
its direction, in contradiction to the isotropy of the physical situation.
For all slip regimes we obtain the same dispersion relation for
two-dimensional interfaces as for the 
corresponding problem with one-dimensional interfaces,
except that the wavenumber is now replaced by the absolute
value of the wave vector.

\section{Spinodal dewetting}
\label{sec:dewet}

\subsection{The no-, weak-, and intermediate slip limit}

If the slip length $b$ is small compared to the lateral length
scale $L$ in the dewetting film (i.e., the spinodal wavelength, see
below), or comparable to $L$, the dynamics of a thin non-volatile
Newtonian liquid
film between a vapour of negligible viscosity and density, and an
impermeable substrate is given in the lubrication
approximation (i.e., for $\varepsilon=H/L\ll 1$, with the mean film
thickness $H$) by a degenerate parabolic partial
differential equation of fourth order for the film thickness
$h(x,t)$ \cite{oron97}
\begin{equation}
\label{weakslip}
\partial_t h= -\partial_x\left\{ M(h) \partial_x \left[
\Pi(h)+\sigma\,\partial^2_x h\right]\right\}\,
\end{equation}
with the surface tension of the
liquid-vapour interface $\sigma$, and the disjoining pressure (DJP) 
$\Pi(h)=-\partial_h \Phi(h)$ (the negative derivative with
respect to the film thickness of the effective interface
potential) \cite{degennes85,dietrich88}. 
If the slip length is on the order of 
$H$ or smaller, the so-called weak-slip regime, 
the mobility factor is given by
$M(h)=\left(h^3/3+b\,h^2\right)/\eta$, 
with the fluid viscosity $\eta$. 
For $b\sim L$, we have $M(h)=b\,h^2/\eta$, called the intermediate-slip regime. 
The well-known no-slip regime is reached
by taking the limit $b\to 0$ in the weak-slip regime, leading to
$M(h)=h^3/(3\,\eta)$, see \cite{witelski05} for more details. 

A homogeneous flat film of thickness $H$ is linearly unstable
if $\partial_h^2 \omega(H)=-\partial_h\Pi(H)<0$\/: In the early
regime of dewetting, we can linearize Eq.~\eqref{weakslip} about the
base state $H$\/. For small perturbations $\delh(x,t)=H-h(x,t)$ we
get 
\begin{equation}
\label{weaklin}
\partial_t \delh = -M(H)\,\partial_x^2\left[
\partial_h\Pi(H)+\sigma\,\partial_x^2\delh\right].
\end{equation}
The ansatz $\delh(x,t)=\delh(q,t)\,\exp(i\,q\,x)$ corresponds to a 
Fourier transformation with respect to $x$ and leads to solutions
of the form $\delh(q,t)=\delh(q,0)\,\exp\left[\omega(q)\,t \right]$ with
the dispersion relation 
\begin{align}
\label{disprel}
\omega(k)&=
M(H)\,q^2\,\left[\partial_h\Pi(H)-\sigma\,q^2\right]\nonumber\\
&= \frac{1}{T}\,(q/Q)^2\,\left[2-(q/Q)^2\right].
\end{align}
For unstable films, i.e., for $\partial_h\Pi(H)>0$, there is a band
of unstable modes with wave number $0<q<q_c$, with
$q_c=\sqrt{[\partial_h\Pi(H)/\sigma}$\/. The dispersion relation
has a maximum at $Q=q_c/\sqrt{2}$, which also defines the 
typical lateral length scale $L=2\,\pi/Q$ which is also called the
spinodal wavelength.
The typical time scale, i.e., the inverse growth rate of the
fastest growing mode, is given by
$T=1/\left[\sigma\,M(H)\,Q^4\right]$\/. 

The dispersion relation $\omega(q)$ has the form given in
Eq.~\eqref{disprel} in the no-slip regime, in the weak-slip, as
well as in the intermediate-slip regime\/. 
For $b \lesssim L$ changing the slip length therefore only
changes the time scale $T$ but not the position $Q$ of the maximum
of $\omega(q)$\/. $Q$, on the other hand, is only determined by
the ratio of $\partial_h\Pi(H)$ and $\sigma$\/. If one knows
$\sigma$, measuring the position of the maximum of $\omega(q)$
for a number of film thicknesses allows to determine
$\partial_h\Pi(h)$ and therefore the effective interface potential
$\Phi(h)$ \cite{seemann01d,seemann01b}\/.
This has been accomplished experimentally by measuring the power spectrum of the surface
roughness $S(q,t)= |\delh(q,t)|^2$, which, in turn, can be calculated from the
initial spectrum $S(q,0)= |\delh(q,0)|^2$ and the dispersion
relation $\omega(q)$
\begin{equation}
\label{power}
S(q,t)=S(q,0)\,\exp\left[2\,\omega(q)\,t\right]\/.
\end{equation}
If $S(q,0)$ is flat in the range of unstable modes, then
$S(q,t)$ has a maximum at the same position as 
$\omega(q)$, i.e., at $q=Q$\/.

\subsection{Strong-slip limit}

For the case of a slip length $b$ much larger than $L$, 
the thin film evolution can be captured by a different 
thin film model \cite{witelski05,kargupta04}, 
called the strong-slip model by M\"unch et al. \cite{witelski05}\/.  
It can be written as 
\begin{subequations}
\label{sslip}
\begin{align}
\label{sslipu}
\eta\,u=&
4\,b\,\eta\,\partial_x\left(h\, \partial_x u\right) + 
b\,h\,\partial_x\left[ \Pi(h)+\partial_x^2 h\right] \nonumber\\
&-
b\,h\,\rho\,\left(\partial_t u + u\,\partial_x u \right) \\
\label{ssliph}
\partial_t h=& -\partial_x\left(h\,u\right),
\end{align}
\end{subequations}
with the fluid mass density $\rho$ and the horizontal flow
velocity $u(x,t)$\/. 
We note that this model is associated with plug flow in the cross-section.
For the 
experimental systems considered here we are not interested in the last term in
Eq.~\eqref{sslipu} since the inertial term proportional to $\rho$ 
is negligible. However, for completeness, we calculate the
dispersion relation including this term. The first term on the right hand side of
Eq.~\eqref{sslipu} is proportional to the divergence of the total
longitudinal shear stress component parallel to the substrate.
Note that the velocity $u$ cannot be eliminated from
Eq.~\eqref{sslip} even if the inertial term is neglected. 

If we perturb Eq.~\eqref{sslip} about a resting [$u(x,0)=0$]
flat film of thickness $H$
we get to first order in the perturbation the problem 
\begin{subequations}
\label{lsslip}
\begin{align}
\label{lsslipu}
\eta\,\delu=&
4\,b\,H\,\eta\,\partial_x^2\delu + 
b\,H\,\partial_x\left[
\partial_h\Pi(H)\,\delh+\partial_x^2\delh\right] \nonumber\\
&-
b\,H\,\rho\,\partial_t \delu  \\
\label{lssliph}
\partial_t\delh=& -H\,\partial_x \delu.
\end{align}
\end{subequations}
With the normal modes ansatz
$\delh(q,t)=\delh(q,0)\,\exp\left[\omega(q)\,t+i\,q\,x\right]$ and
$\delu(q,t)=\delu(q,0)\,\exp\left[\omega(q)\,t+i\,q\,x\right]$ we
get, after taking the derivative of Eq.~\eqref{lssliph} with respect
to $x$ and subsequently eliminating $\delh(q,0)$, a quadratic equation for
the dispersion relation $\omega(q)$
\begin{align}
\label{cheq}
\eta\,\omega(q)=&
-4\,b\,H\,\eta\,q^2\,\omega(q)
+b\,H^2\,q^2\,\left[\partial_h\Pi(H) -\sigma\,q^2\right]\nonumber\\
&-b\,H\,\rho\,\left[\omega(q)\right]^2
\end{align}
with the two solutions
\begin{multline}
\label{ssdisprel}
\omega_{1/2}(q)= -\frac{\eta}{2\,\rho}\,
\left(4\,q^2+\frac{1}{b\,H}\right)\\
\times \left[
1\pm\sqrt{
1+\frac{H\,q^2\,\left[\partial_h\Pi(H)-\sigma\,q^2\right]}{\frac{\eta^2}{\rho}\,\left(4\,q^2+\frac{1}{b\,H}\right)}
}
\right].
\end{multline}
While the first solution
$\omega_1(q)$, corresponding to the plus sign in
Eq.~\eqref{ssdisprel}, is negative for all $q$ and does therefore
not contribute significantly to the roughness spectrum, the 
second solution $\omega_2(q)$ is zero for
$q^2\,[\partial_h\Pi(H)-\sigma\,q^2]=0$, i.e., for
$q=0$ and for $q=q_c=\sqrt{2}\,Q$\/. For sufficiently small $\rho$
or sufficiently large $b$ the term under the square root is negative
for large enough $q>q_c$\/. The corresponding modes 
oscillate in time. However, there real part is given by
$\text{Re}\,\omega_2(q)=-\eta\,\left[4\,q^2+1/(b\,H)\right]/(2\,\rho)<0$
and thus, these modes are heavily damped. 

\begin{figure}
\includegraphics[width=\linewidth]{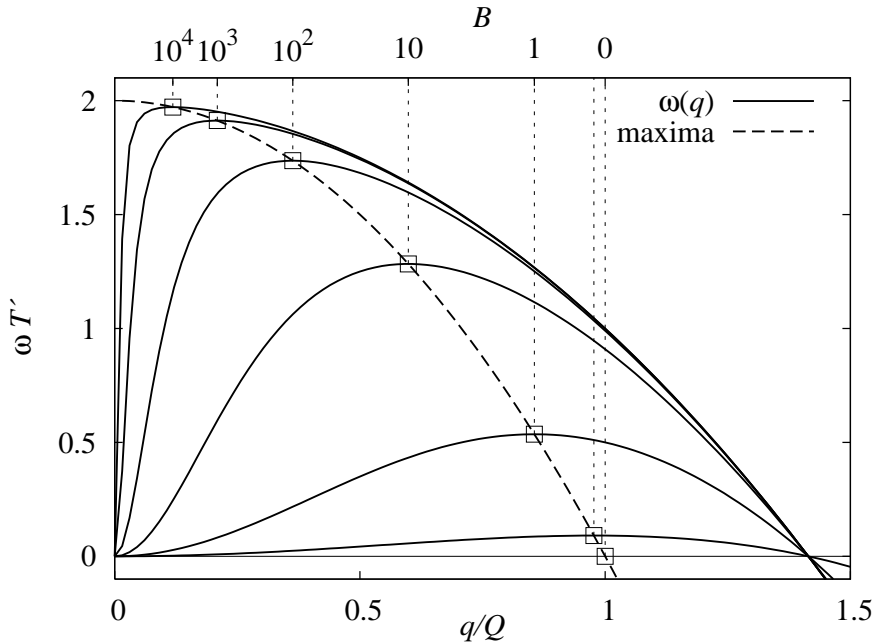}
\caption{\label{disprelfig} Dispersion relation
in the strong slip limit in Eq.~\protect\eqref{sssdisprel} (full
lines) vor various values of $B$ (see upper axis tics)\/.
The location of the 
maximum of the dispersion relation shifts to smaller values $q/Q$ and the
height approaches $2/T'$ from below for increasing $B$ (dashed line)\/.}
\end{figure}

If inertia is negligible we can ignore the last term proportional
to $\rho$ in Eq.~\eqref{cheq} (or take the limit $\rho\to 0$ 
in Eq.~\eqref{cheq}) and we get
\begin{align}
\label{sssdisprel}
\omega(q)&=\frac{b\,H^2\,q^2\,\left[\partial_h\Pi(H)-\sigma\,q^2\right]}
{\eta\,\left(1+4\,b\,H\,q^2\right)}\nonumber\\
&=\frac{(q/Q)^2\,\left[2-(q/Q)^2\right]}
{T\,\left[1+B\,(q/Q)^2\right]},
\end{align}
with the time scale $T=\eta/\left(\sigma\,b\,H^2\,Q^4\right)$,
as in the inter\-mediate-slip regime. 
As expected, taking the limit $B\to 0$ ($b\to 0$)
we recover the dispersion relation of the intermediate-slip model.
As compared to Eq.~\eqref{disprel}, we have one additional
dimensionless parameter $B=4\,b\,H\,Q^2$ determining the relevance of
slip, which is of order unity in the strong-slip limit discussed here. 
Both the time scale $T$ as well as $B$ depend on the
slip length. Introducing the $b$-independent time scale
$T'=B\,T=4\,\eta/(\sigma\,H\,Q^2)$ we can study the dependence of
the dispersion relation on the slip length more easily. In the
strong-slip model, neither $T$ nor $T'$ are the inverse of the
maximum of the dispersion relation.  Since the numerator in
Eq.~\eqref{sssdisprel} is proportional to the dispersion relation
of the weak- and no-slip models discussed in the previous section and
since the denominator is positive, the strong slip model has the
same band of unstable modes $q<q_c$\/.
Taking the derivative of Eq.~\eqref{sssdisprel} we get the
position $q_{\text{max}}$ of the maximum of $\omega(q)$ at
\begin{equation}
\label{peakpos}
q_{\text{max}}=Q\,\sqrt{\frac{\sqrt{1+2\,B}-1}{B}}.
\end{equation}
Fig.~\ref{disprelfig} shows the dispersion relation for the strong-slip 
regime for various values of $B$\/.
For $B\to 0$ we recover the shape of the dispersion relation for the
weak and intermediate slip regime, but the time scale $T$ diverges
in this limit. For increasing $B$ the location $q_{\text{max}}$ of
the maximum moves to smaller values of $q$  and the height of the
maximum $\omega(q_{\text{max}})$ approaches $2/T'$ from below. In
the limit $B\to\infty$ we get $\omega(q)\to
\left[2-(q/Q)^2\right]/T'$\/.

\section{Experimental relevance}
\label{sec:discussion}

In the experiments discussed in \cite{seemann01d} the shift of $Q$
with the film thickness was used to determine the effective
interface potential $\Phi(z)$, assuming the dispersion relation in
Eq.~\eqref{disprel}, i.e., for the weak/intermediate slip regime. 
The surface tension coefficient was
$\sigma=31$~mN/m and the DJP had the form
$\Pi(z)=-A/\left(6\,\pi\,z^3\right)$, with the Hamaker constant
$2.2\times 10^{-20}$~Nm\/. From this we get for the dimensionless
slip length $B=0.23\,\text{nm}^2\,b/H^3$\/. Therefore, in
order to have, e.g., $B>1$ or $B>0.1$ for the lowest film thickness
$H=2$~nm in the experiment, $b$ has to be larger than 35~nm and
3.5~nm, respectively.

\begin{figure}
\includegraphics[width=\linewidth]{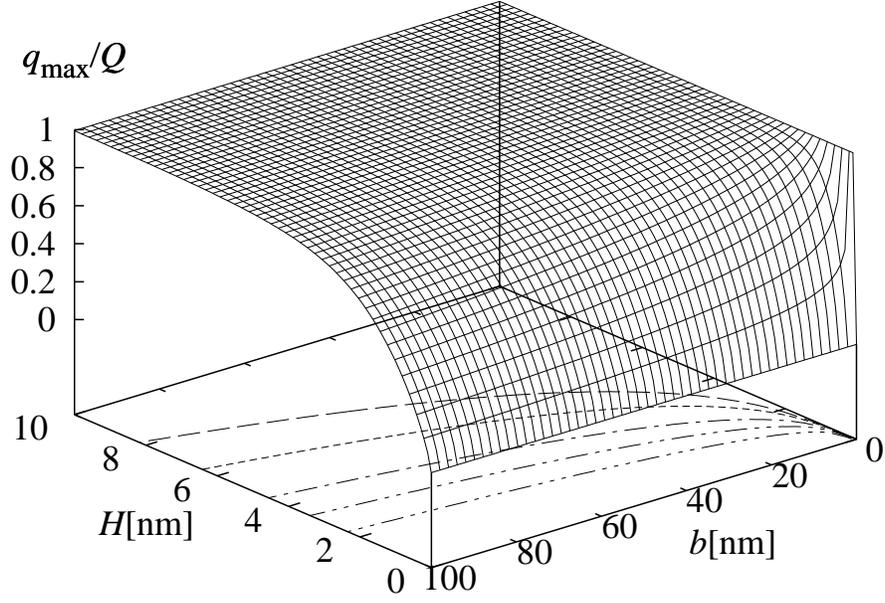}\\
\includegraphics[width=\linewidth]{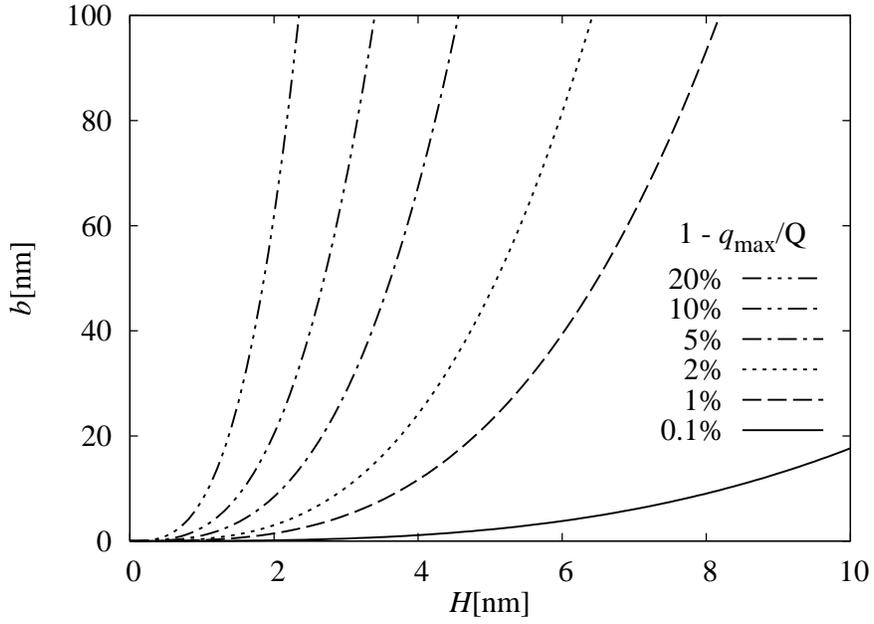}
\caption{\label{expfig} Shift of the position of the maximum
$q_{\text{max}}$ of the
dispersion relation $\omega(q)$ as compared to the position
expected for the weak and intermediate slip model and experimental
parameters from \protect\cite{seemann01d} as a function of
the film thickness $H$ and the slip length $b$\/. The contour lines
indicate the slip length $b$ necessary in order to obtain a
deviation of 0.1\%\/, 1\%\/, 5\%\/, 10\%\/, and 20\%\/.}
\end{figure}

The position of the peak in the power spectrum as a function of $H$
and $b$ normalized to the position in the weak/intermediate slip
model that one would
observe for the material combination studied in \cite{seemann01d}
is shown in Fig.~\ref{expfig}\/. Cleary, the shift
in the peak is larger for smaller film thicknesses $H$ and larger
slip length $b$\/. In order to get a deviation of the peak position
on the order of 5\%\ for the smallest film thicknesses of $H=2$~nm,
the slip length has to be larger than 8~nm, i.e., much larger than
expected for PS on Si\/.

If, on the other hand, the strong-slip regime was to apply and one
tried to determine $\partial_h \Pi$ from the measured peak position 
$q_{\text{max}}$  [see Eq.~\eqref{peakpos}] with the equation valid
only in the weak and intermediate slip regime, i.e.,
$q_{\text{max}}=\sqrt{\partial_h\Pi^*/\left(2\,\sigma\right)}$
with an ``apparent'' DJP $\Pi^*$, one would 
produce a systematic error in the measurment. The ratio of the
actual DJP $\partial_h\Pi$ and the ``apparent'' DJP $\partial_h\Pi^*$ is
obtained by squaring Eq.~\eqref{peakpos}
\begin{equation}
\partial_h\Pi^*(H)=\frac{\sqrt{1+2\,B}-1}{B}\,\partial_h\Pi(H).
\end{equation}
In order to get a first estimate on the error we assume that only
non-retarded dispersion forces are relevant and get to first order
in $B$
\begin{equation}
\partial_h\Pi^*(H) \approx \frac{A}{2\,\pi\,H^4}
-\frac{A^2\,b}{4\,\pi^2\,\sigma\,H^7}.
\end{equation}
Therefore, a spurious subleading term $\propto 1/z^5$ is
generated in the ``apparent'' effective interface potential
$\Phi^*(z)=-\int^z \Pi^*(z')\,dz'$\/. 

\begin{figure}
\includegraphics[width=\linewidth]{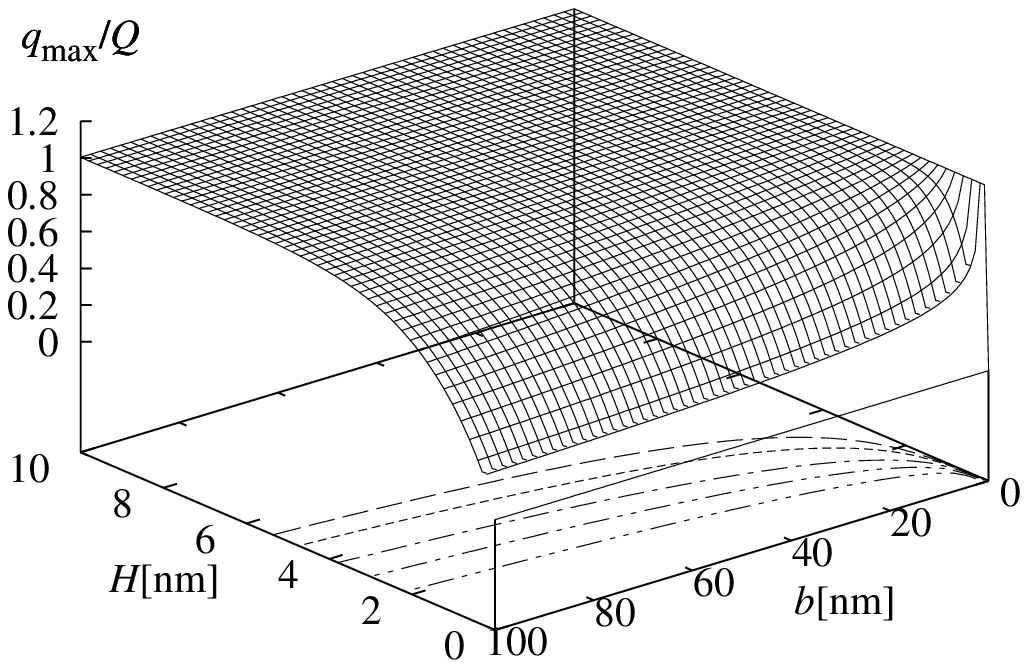}\\
\includegraphics[width=\linewidth]{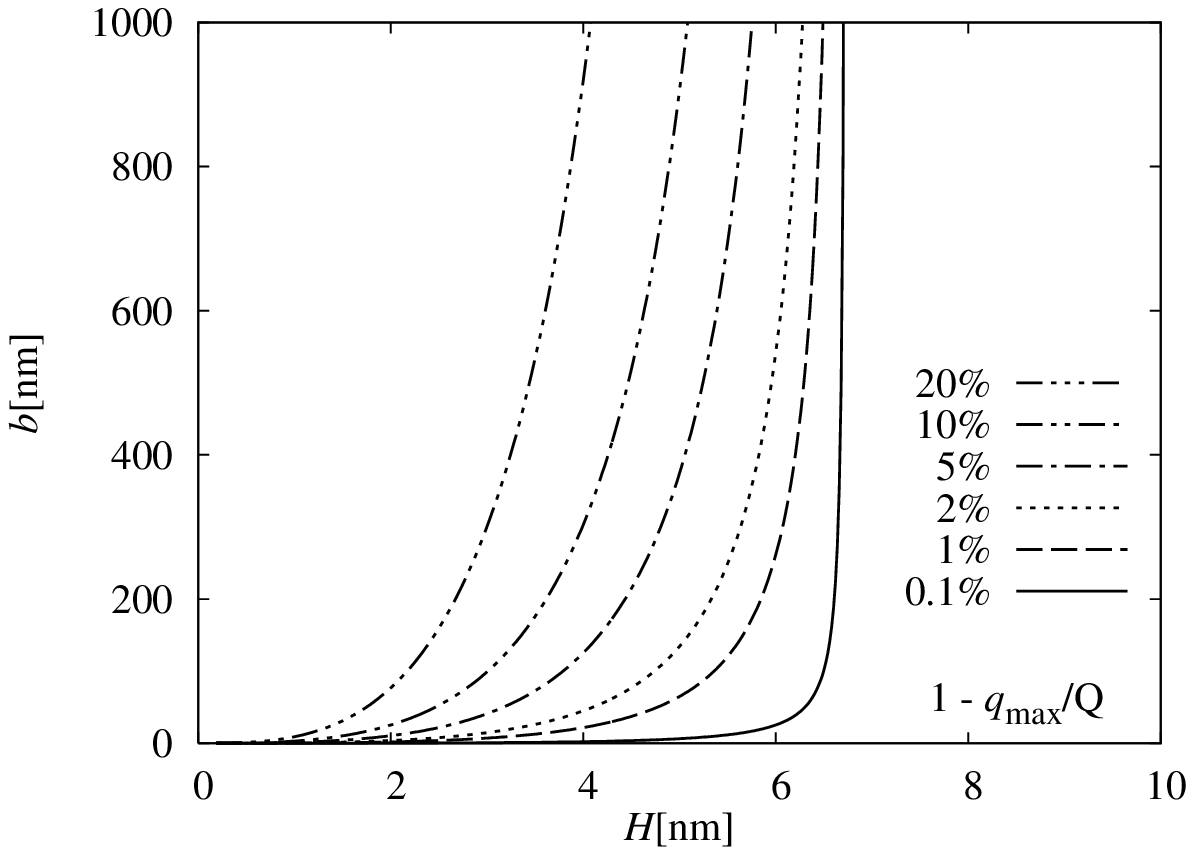}
\caption{\label{OTSfig} Shift of the position of the maximum
$q_{\text{max}}$ of the
dispersion relation $\omega(q)$ as compared to the position
expected for the weak and intermediate slip model and experimental
parameters for a PS film on OTS from  \protect\cite{seemann01b} as a function of
the film thickness $H$ and the slip length $b$\/. The contour lines
indicate the slip length $b$ necessary in order to obtain a
deviation of 0.1\%\/, 1\%\/, 5\%\/, 10\%\/, and 20\%\/.}
\end{figure}

The system considered in \cite{seemann01d}, i.e., PS on a Si wafer
covered with a native oxide layer, is known not to exhibit
significant slip. However, recently it has been demonstrated, that
covering the same wafer with an OTS or DTS brush leads to very large slip
lengths up to the order of microns
\cite{fetzer05,fetzer06a,fetzer07a}\/. 
With the material parameters
of OTS, SiO, and Si together with the thickness of the OTS layer
and of the SiO layer in \cite{fetzer05} we calculate the effective
interface potential for a PS film on an OTS covered Si waver using Eq.~(3) in
Ref.~\cite{seemann01b}\/. With this, we can calculate the
deviation of the position of the maximum of the dispersion relation
$q_{\max}$ from the position in the weak-slip limit $Q$ as shown in
Fig.~\ref{OTSfig}\/. With a
film thickness of 4~nm a slip length of $b=1\,\mu$m is enough to
generate a 20\%\ shift in the maximum of the dispersion relation.
Such a large shift should be detectable in the experiments.

\section{Conclusions}
\label{sec:summary}

 In this paper we demonstrated that the hydrodynamic boundary condition at the
substrate surface significantly changes the power spectrum of
film thickness variations in spinodal dewetting for
experimentally relevant systems.
Analysing only the peak position without knowledge of the
hydrodynamic boundary conditions can lead to significant systematic
errors in the data analysis. 
As pointed out in \cite{blossey06a},
viscoelastic thin films show a similar behaviour:
while the position of the
maximum of the dispersion relation is identical to the position in
the Newtonian weak-/intermediate-slip case, in the strong-slip case 
it shifts to smaller wave numbers for increasing slip length
\cite{rauscher05a,blossey06a,muench06a}\/.

The power spectrum of capillary waves should be affected by
hydrodynamic slip as well.
However, up to now, a stochastic version of the thin-film
equation is available only for substrates without slip
\cite{mecke05a,gruen06a} and the phenomenological ansatz taken in
\cite{davidovitch05} can be extended directly to the case of weak- and
intermediate-slip only. In the case of vanishing slip the 
position of the maximum of the power spectrum approaches $Q$
from above as time proceeds \cite{fetzer07b}\/. Since the
no-slip, the weak-slip, and the intermediate-slip case differ
only in the mobility factor $M$, the same
behaviour can be expected for the weak and intermediate-slip
case. The mechanism for this noise-induced coarsening is 
simple: thermal fluctuations
generate short wavelength fluctuations rather rapidly before
the instability sets in, amplifying modes with larger
wavelength. In the strong slip case, the maximum of the
dispersion relation shifts to very large wavelengths for
increasing $b$, which could emphasize the effect of noise-induced
coarsening. However, a detailed analysis of a
stochastic strong-slip thin-film equation is needed to reach a
conclusion on this point.

\section*{Acknowlegement}
The authors thank K. Jacobs for inspiring
discussions.
A.M., B.W., and M.R. acknowledge funding from DFG priority program SPP 1164
``Nano- and Microfluidics''.

\begin{appendix}
\section*{Appendix}

The generalization of the strong-slip model to 3D has been derived in 
\cite{ambw08}\/. With the two lateral velocity components $u$ and $v$  
in the $x$ and $y$-direction, respectively, the model equations in
non-dimensional form are
\begin{subequations}
\label{ssthreedee}
\begin{align}
\label{ssthreedeeu}
\mbox{Re}\frac{du}{dt}=& \frac{1}{h}\left[\partial_x\left(4 h\partial_x u + 2h\partial_y v\right) +\partial_y\left(h\partial_x v + h\partial_y u\right)\right] \nonumber\\
             & +\partial_x\left[\Delta h +\Pi(h)\right] -\frac{u}{h\beta}\\
\label{ssthreedeev}
\mbox{Re}\frac{dv}{dt}=& \frac{1}{h}\left[\partial_y\left(4 h\partial_y v + 2h\partial_x u\right) +\partial_x\left(h\partial_x v + h\partial_y u\right)\right] \nonumber\\
             & +\partial_y\left[\Delta h +\Pi(h)\right] -\frac{v}{h\beta} \\
\label{ssthreedeeh}
\partial_t h=& -\partial_x\left(h\,u\right) -\partial_y\left(h\,v\right).
\end{align}
\end{subequations}
Here, we abbreviate the total/materials derivative $d/dt=\partial_t
+u\,\partial_x + v\,\partial_y$ and the two-dimensional Laplace
operator $\Delta = \partial_x^2+\partial_y^2$\/ and 
Re is the Reynolds number. The lateral
length scale $L$, the vertical length scale $H$, and the time scale
$T$ have been introduced in the main text. For this model the slip length 
is large and of order $b=\beta/\varepsilon^2$, where 
$\beta$ is an $O(1)$ constant. The scale for the
(disjoining) pressure is $P=\eta/T$\/. The capillary number is
$\text{Ca}=\eta\,L/(\sigma\,T)= \varepsilon$\/. 

The linear stability of a flat film is again a straightforward calculation, 
by taking the first order in the perturbation, the problem for 
$u=\delta\!u$,  $v=\delta\!v$ and $h=H+\delta\!h$ and making the
normal modes ansatz $(\delta\!u, \delta\!v, \delta\!h)(\vec{q},t) =
(\delta\!u(\vec{q},0), \delta\!v(\vec{q},0), \delta\!h(\vec{q},0))\,
\exp[\omega(\vec{q})\,t + i\,\vec{q}\cdot\vec{r}]$, with
$\vec{q}=(q_x,q_y)$ and $\vec{r}=(x,y)$\/. 

Abbreviating 
$\delta\!u(\vec{q},0)=\delta\!u_0$, $\delta\!v(\vec{q},0)=\delta\!v_0$,  
$\delta\!h(\vec{q},0)=\delta\!h_0$, 
we obtain the linear eigenvalue problem
\begin{subequations}
\label{linthreedee}
\begin{align}
\label{linthreedeeu}
\eta\,\delta\!u_0 =&
-b\,\eta\,H\,\left[(4\,q_x^2+q_y^2)\,\delta\!u_0 +
3\,q_x\,q_y\,\delta\!v_0\right] \nonumber \\ &
+i\,b\,H\,q_x\,\left[\partial_H\Pi(H)-\sigma\,q^2\right]\,\delta\!h_0\\
\label{linthreedeev}
\eta\,\delta\!v_0 =&
-b\,\eta\,H\,\left[(4\,q_y^2+q_x^2)\,\delta\!v_0 +
3\,q_x\,q_y\,\delta\!u_0\right] \nonumber \\ &
+i\,b\,H\,q_y\,\left[\partial_H\Pi(H)-\sigma\,q^2\right]\,\delta\!h_0\\
\label{linthreedeeh}
\omega(\vec{q})\,\delta\!h_0 =&
-i\,H\,(q_x\,\delta\!u_0+q_y\,\delta\!v_0),
\end{align}
\end{subequations}
with $q^2=|\vec{q}|^2$\/. Here, we have neglected the
contributions from the inertial terms and we have switched to
dimensional quantities in order to connect to the main body of the
article. From \eqref{linthreedee} we
find the same dispersion relation as we obtained for the 2D case in
\eqref{sssdisprel}, i.e., $\omega(\vec{q})=\omega(q)$, by setting
the determinant of the matrix corresponding to the linear system to
zero and by solving for $\omega(q)$\/. Alternatively, one can
determine $q_x\,\delta\!u_0+q_y\,\delta\!v_0$  from Eqs.~\eqref{linthreedeeu}
and \eqref{linthreedeev} which yields 
\begin{multline}
(q_x\,\delta\!u_0+q_y\,\delta\!v_0)\,(1 + 4\,b\,H\,q^2) =\\
\frac{i\,b\,H}{\eta}\,q^2\,\left[\partial_H\Pi(H) -
\sigma\,q^2\right]\,\delta h_0\,.  
\end{multline}
Inserting this in Eq.~\eqref{linthreedeeh} we recover
Eq.~\eqref{sssdisprel}\/.
 
\end{appendix}


\providecommand{\url}[1]{\texttt{#1}}
\providecommand{\refin}[1]{\\ \textbf{Referenced in:} #1}

\end{document}